\documentclass[twocolumn,showpacs,prl,amsmath,amssymb]{revtex4}
\usepackage[english]{babel}
\usepackage{latexsym}
\usepackage{graphicx}
\usepackage{subfigure}

\begin{document}

\title{Observation of Electromagnetically Induced Transparency and Slow Light in the Dark State - Bright State Basis of CPT}

\author{T. Laupr\^etre$^1$, J. Ruggiero$^1$, R. Ghosh$^2$, F. Bretenaker$^1$ and F. Goldfarb$^1$}
\email{Fabienne.Goldfarb@lac.u-psud.fr}

\affiliation{$^1$ Laboratoire Aim\'e Cotton, CNRS-Universit\'e Paris Sud 11, 91405 Orsay Cedex, France
\\
$^2$ School of Physical Sciences, Jawaharlal Nehru University, New Delhi 110067, India
}

\date{\today}


\begin{abstract}
Electromagnetically induced transparency (EIT) is observed in a three-level system composed of an excited state and two coherent superpositions of the two ground-state levels. This peculiar ground state basis is composed of the so-called bright and dark states of the same atomic system in a standard coherent population trapping configuration. The characteristics of EIT, namely, width of the transmission window and reduced group velocity of light, in this unusual basis, are theoretically and experimentally investigated and are shown to be essentially identical to those of standard EIT in the same system.
\end{abstract}

\maketitle
Electromagnetically induced transparency (EIT) in three level $\Lambda$-systems is based on quantum interference effects involving  coherence between the two lower levels \cite{Harris1990,Boller1991,Scully1997}. This effect has raised a great interest because it opens the door to many fascinating phenomena such as slow light \cite{Harris1992,Hau1999}, stopped light \cite{Kocharovskaya2001,Liu2001,Phillips2001}, or enhancement of nonlinear effects \cite{Harris1999,Kash1999,Wang2001}. EIT is usually interpreted as being due to the fact that the atoms are pumped into a so-called dark state \cite{Arimondo1976,Alzetta1976} which is not coupled to the excited state. EIT is then considered as a limit case of Coherent Population Trapping (CPT) \cite{Arimondo1996} in which the probe field Rabi frequency is vanishingly small compared with the Rabi frequency of the coupling field \cite{Fleischhauer2005}. In this case, the dark and bright states are simply the initial two ground-state levels. \\
The aim of the present Letter is to show that EIT can be observed for any superposition of the two ground-state levels of the $\Lambda$- system. In particular, we investigate experimentally whether EIT can be obtained when the starting ground-state levels are the half/half superpositions of the usual ground-state levels, i.\,e., the so-called dark and bright states for the same atom in a symmetrical CPT situation \cite{Arimondo1996,Aspect1988}.

\begin{figure}
\begin{center}
\includegraphics[width=8.0 cm]{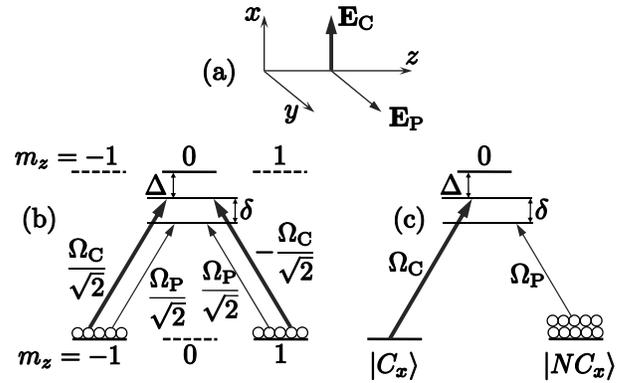}
\end{center}
\caption{(a) Polarizations of the coupling and probe beams, both propagating along $z$. (b) Excitation scheme in the usual basis with the quantization axis along $z$. (c) Excitation scheme when the two ground states are described in the dark state-bright state basis.}\label{Fig01}
\end{figure}
The level scheme relevant to the present experiment is described in Fig.\,\ref{Fig01}. As can be seen in Fig.\,\ref{Fig01}(a), we consider two beams propagating along $z$: an intense coupling beam linearly polarized along $x$ and a weak probe beam linearly polarized along $y$. We suppose that these two beams are quasi-resonant with a $J=1\rightarrow J=1$ atomic system, as shown in Fig.\,\ref{Fig01}(b). We denote by $\Delta$ the optical detuning of the coupling beam and by $\Delta+\delta$ the detuning of the weak probe beam.  With the quantization axis along $z$, the experiment that we consider here looks like a ``double CPT'' scheme. Indeed, we can see in Fig.\,\ref{Fig01}(b) that since the $x$ and $y$ linear polarizations can be decomposed equally into a $\sigma^+$ and a $\sigma^-$ component according to:
\begin{subequations}
\label{eqTas1}
\begin{equation}
|\hat{\mathbf{x}}\rangle=\frac{1}{\sqrt{2}}\left(|\mathbf{\hat{\sigma}^+}\rangle+|\mathbf{\hat{\sigma}^-}\rangle\right)\ ,\label{eq1}
\end{equation}
\begin{equation}
|\hat{\mathbf{y}}\rangle=\frac{-\mathrm{i}}{\sqrt{2}}\left(|\mathbf{\hat{\sigma}^+}\rangle-|\mathbf{\hat{\sigma}^-}\rangle\right)\ , \label{eq2}
\end{equation}
\end{subequations}
%\begin{eqnarray}
%|\hat{\mathbf{x}}\rangle&=&\frac{1}{\sqrt{2}}\left(|\mathbf{\hat{\sigma}^+}\rangle+|\mathbf{\hat{\sigma}^-}\rangle\right)\ ,\label{eq1}\\
%|\hat{\mathbf{y}}\rangle&=&\frac{-\mathrm{i}}{\sqrt{2}}\left(|\mathbf{\hat{\sigma}^+}\rangle-|\mathbf{\hat{\sigma}^-}\rangle\right)\ , \label{eq2}
%\end{eqnarray}
each beam excites the two legs of the $\Lambda$ system equally. We call $\Omega_{\mathrm{C}}$ ($\Omega_{\mathrm{P}}$) the Rabi frequency which would correspond to the coupling (probe) beam intensity if it were circularly polarized, as in a standard EIT configuration in a $J=1\rightarrow J=1$ transition. Then the different Rabi frequencies are as given in Fig.\,\ref{Fig01}(b). We can see that since the $m_z=0\rightarrow m_z=0$ transition is forbidden in a $J=1\rightarrow J=1$ system, the system is reduced to a pure three-level $\Lambda$ system after a few optical pumping cycles. However, the fact that both beams excite both legs of the $\Lambda$ system of Fig.\,\ref{Fig01}(b) makes the situation apparently complicated. This is no longer the case if one changes the basis from the two lower levels $|-1\rangle$ and $|+1\rangle$ to the following one:
\begin{subequations}
\label{eqTas2}
\begin{equation}
|C_x\rangle=|NC_y\rangle=\frac{1}{\sqrt{2}}\left(|-1\rangle-|+1\rangle\right)\ ,\label{eq3}
\end{equation}
\begin{equation}
|NC_x\rangle=|C_y\rangle=\frac{1}{\sqrt{2}}\left(|-1\rangle+|+1\rangle\right)\ ,\label{eq4}
\end{equation}
\end{subequations}
where the states $|C_x\rangle$ and $|C_y\rangle$ ($|NC_x\rangle$ and $|NC_y\rangle$) are the bright (dark) states for a CPT experiment that would be performed with a $x$ or $y$ polarized beam only. In this basis {[see Fig.\,\ref{Fig01}(c)]}, since, using Eqs.\,(\ref{eqTas1}), the coupling (probe) beam couples the excited level to the $|C_x\rangle$ ($|NC_x\rangle$) level only, we can see that we are dealing with a simple EIT configuration in a new basis defined by the coupling laser field. However, the question that remains to be answered is: since our ground-state levels are coherent superpositions of the ``usual'' levels $|-1\rangle$ and $|+1\rangle$, can we observe the usual EIT effect in such a configuration?

To answer this question, we have used a well-known closed $J=1\rightarrow J=1$ system, namely the $^3$S$_1 \rightarrow ^3$P$_1$ transition of $^4$He, which had already been proved to exhibit CPT \cite{Gilles2001} and EIT \cite{Goldfarb2008} resonances at room temperature. This system is well suited to investigate the present problem because it can be reduced to a pure three-level system, contrary to alkali atoms. In a standard EIT configuration where the coupling beam is $\sigma^+$-polarized and the probe beam is $\sigma^-$-polarized, we have recently observed that EIT in this inhomogeneously broadened system is very well modeled by considering it as an homogeneously broadened system with an effective width $W$ \cite{Goldfarb2008,Ghosh2009,Goldfarb2009}. This effective width is the spectral width over which the atoms are efficiently optically pumped into the $|+1\rangle$ level by the coupling beam with the help of velocity changing collisions \cite{Ghosh2009,Goldfarb2009} and can thus take part in EIT. With this model, the width (full width at half maximum) of the EIT peak at optical resonance ($\Delta=0$) has been shown to be given by:
\begin{equation}
\Gamma_{\mathrm{EIT}}=2\Gamma_{\mathrm{R}}+\frac{\Omega_{\mathrm{C}}^2}{2W+\Gamma}\ , \label{eq5}
\end{equation}
where $\Gamma_{\mathrm{R}}$ and $\Gamma$ are the decay rates of the Raman and optical coherences, respectively. To compare the ``standard'' EIT configuration (with a $\sigma^+$-polarized coupling beam and a $\sigma^-$-polarized probe beam, see Ref.\,\cite{Goldfarb2008}) with the present configuration (see Fig.\,\ref{Fig01}(b)), we have already seen that the relevant Rabi frequency is the same $\Omega_\mathrm{C}$, which is related to the coupling beam intensity in the standard way. We can also reasonably argue that the decay rate of the optical coherences $\Gamma$ and the effective atomic pumping width $W$ will be the same for the two configurations. The only question that remains is to know whether the decay rate of the Raman coherences is the same when one considers the $|-1\rangle$ and $|+1\rangle$ levels or the $|C_x\rangle$ and $|NC_x\rangle$ levels. The relaxation terms that play a role in the density matrix master equation written in the $\left\{|-1\rangle,|+1\rangle\right\}$ basis are \cite{Ghosh2009}:
\begin{subequations}
\label{eqTas3}
\begin{equation}
\left.\frac{\mathrm{d}\rho_{+1,+1}}{\mathrm{d}t}\right|_{relax}=\frac{\Gamma_0}{2}\rho_{00}-\Gamma_{\mathrm{t}}\left(\rho_{+1,+1}-\frac{1}{2}\right)\ ,\label{eq6}
\end{equation}
\begin{equation}
\left.\frac{\mathrm{d}\rho_{-1,-1}}{\mathrm{d}t}\right|_{relax}=\frac{\Gamma_0}{2}\rho_{00}-\Gamma_{\mathrm{t}}\left(\rho_{-1,-1}-\frac{1}{2}\right)\ ,\label{eq7}
\end{equation}
\begin{equation}
\left.\frac{\mathrm{d}\rho_{-1,+1}}{\mathrm{d}t}\right|_{relax}=-\Gamma_{\mathrm{R}}\ \rho_{-1,+1}\ ,\label{eq8}
\end{equation}
\end{subequations}
where $\Gamma_0$ is the upper level population lifetime and $\Gamma_{\mathrm{t}}$ is the transit time of the atoms through the beam, which plays the role of an effective lifetime for the ground-state sublevels populations. Using Eqs.\,(\ref{eqTas2}) to change to the $\left\{|C_x\rangle,|NC_x\rangle\right\}$ basis, Eqs.\,(\ref{eqTas3}) lead to:
\begin{equation}
\left.\frac{\mathrm{d}\rho_{C_x,NC_x}}{\mathrm{d}t}\right|_{relax}=-\Gamma_{\mathrm{R}}\ \rho_{C_x,NC_x}-\frac{\Gamma_{\mathrm{t}}}{2}(\rho_{+1,+1}-\rho_{-1,-1})\ .\label{eq9}
\end{equation}
In the configuration of Fig.\,\ref{Fig01}(b), the $|-1\rangle$ and $|+1\rangle$ sublevels play completely symmetric roles, thus one has $\rho_{+1,+1}=\rho_{-1,-1}$, leaving us with:
\begin{equation}
\left.\frac{\mathrm{d}\rho_{C_x,NC_x}}{\mathrm{d}t}\right|_{relax}=-\Gamma_{\mathrm{R}}\ \rho_{C_x,NC_x}\ .\label{eq10}
\end{equation}
Surprisingly, we thus expect the configuration of Fig.\,\ref{Fig01} to lead exactly to the same dependence of the resonance width on the coupling beam intensity.

\begin{figure}
\begin{center}
\includegraphics[width=7 cm]{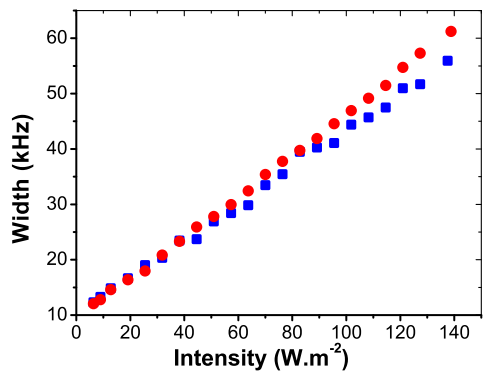}
\end{center}
\caption{Measured evolutions vs the coupling beam intensity of the widths of the EIT peaks in standard EIT configuration (circles) and in the configuration of Fig.\,\ref{Fig01}(b, c) (squares). The linear behavior corresponds to effective pumping half-width of 0.46 and 0.53\,GHz, respectively.}\label{Fig02}
\end{figure}
We check this using a 2.5 cm long $^4$He cell filled at 1 Torr. The $^3\mathrm{S}_1$ metastable level is populated by an RF discharge at 27\ MHz. The cell is located in a three-layer mu-metal shield to reduce spurious magnetic fields. At the center of the Doppler profile of the optical transition and for a vanishing light intensity, the cell absorbs about 60\% of the incident intensity. Light at 1.083 $\mu\mathrm{m}$ is provided by a diode laser (model SDL-6700). The beam is spatially filtered by a single-mode fiber before being separated by a polarizing beamsplitter into a pump beam and a probe beam. Their frequencies and intensities are controlled by two acousto-optic modulators. The pump beam is fed continuously at a fixed frequency. The probe beam can be linearly scanned in frequency or modulated in intensity. These two beams are recombined before entering the cell in a copropagating configuration. For selecting whether we want the pump and probe beams to be linearly or circularly polarized, we just need to introduce a quarter-wave plate in front of the helium cell. The diameter of the Gaussian beam inside the cell is chosen around 1 cm (at $1/\mathrm{e}^2$ of the maximum intensity). After the cell, a polarizer, preceded by a quarter-wave plate in the $\sigma^+/ \sigma^-$ (standard EIT) configuration, permits only the probe beam to reach the detector. In the experiments reported below, the probe laser power is set around 70\,$\mu$W. When the coupling beam frequency is tuned close to the maximum absorption frequency of the $^3\mathrm{S}_1 \rightarrow\,^3\mathrm{P}_1$ transition [$\Delta=0$ in Fig.\,\ref{Fig01}(a)], we probe the EIT window created by this coupling beam by scanning the frequency of the weak probe beam around Raman resonance, thus scanning the Raman detuning $\delta/2\pi$ around 0 between $-$150 and +150 kHz in 3 ms.

The absorption profiles obtained by this method are fitted by Lorentzians. The evolution of the width (full width at half maximum) of these Lorentzians versus coupling beam intensity is reproduced in Fig.\,\ref{Fig02}, for both the ``standard'' $\sigma^+/\sigma^-$ EIT configuration and the $x/y$ configuration of Fig.\,\ref{Fig01}. One can see that the results for the two configurations are almost indistinguishable. A linear fit owing to Eq.\,(\ref{eq5}) leads to a slope equal to 372\,Hz/(W/m$^2$) and an intercept corresponding to $\Gamma_{\mathrm{R}}/2\pi=4.6$\,kHz for the $\sigma^+/\sigma^-$ configuration and to a slope equal to 331\,Hz/(W/m$^2$) and an intercept corresponding to $\Gamma_{\mathrm{R}}/2\pi=5.1$\,kHz for the $x/y$ configuration. The slopes correspond to effective half widths $W/2\pi$ equal to 0.46 and 0.53\,GHz, respectively, which should be compared with the 0.85\,GHz Doppler half width at half maximum.
\begin{figure}
\begin{center}
\includegraphics[width=7 cm]{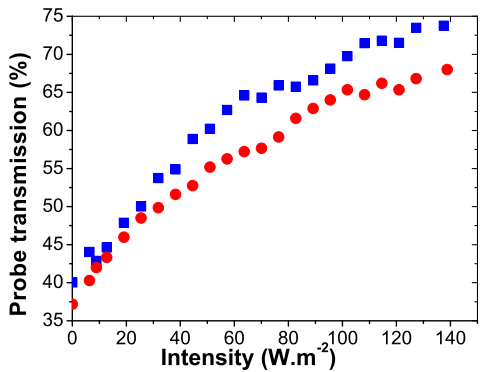}
\end{center}
\caption{Measured evolutions vs the coupling beam intensity of the peak transmission in standard EIT configuration (circles) and in the configuration of Figs.\,\ref{Fig01}(b, c) (squares).}\label{Fig03}
\end{figure}
We have also plotted the probe transmission at the EIT peak ($\Delta=\delta=0$) versus coupling beam power for the two configurations (see Fig.\,\ref{Fig03}). Here again, we can see that the behaviors are similar for the two configurations, as expected and in agreement with our simple model based on an effective width $W$ (for more details see Ref.\,\cite{Goldfarb2009}.)

\begin{figure}
\begin{center}
\includegraphics[width=7 cm]{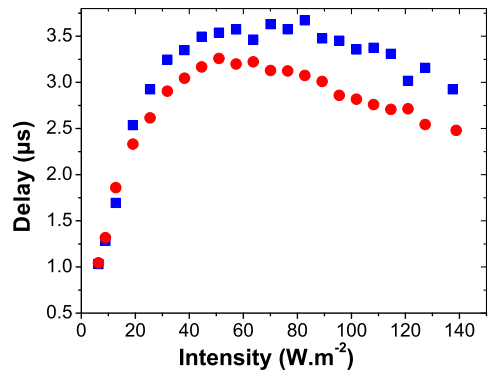}
\end{center}
\caption{Measured evolutions vs the coupling beam intensity of the delay experienced by a 5\,kHz sinusoidal modulation of the probe pulse intensity in standard EIT configuration (circles) and in the configuration of Figs.\,\ref{Fig01}(b, c) (squares).}\label{Fig04}
\end{figure}
Since the results of Figs.\,\ref{Fig02} and \ref{Fig03} indicate that the imaginary part of the susceptibility seen by the probe field is the same in the two configurations, we can expect the real part of this susceptibility to be the same for the $\sigma^+/\sigma^-$ and the $x/y$ configurations. We tested this prediction by measuring the phase shift experienced by a 5\,kHz sinusoidal modulation of a probe pulse versus coupling power for the two configurations. The corresponding results at optical and Raman resonances ($\Delta=\delta=0$) are reproduced in Fig.\,\ref{Fig04}. We can see again that the results obtained for the two configurations are very close, and in agreement with the model of Ref.\,\cite{Goldfarb2009}. The maximum delay is slightly larger for the $x/y$ configuration than for the $\sigma^+/\sigma^-$ configuration. This is consistent with the fact that the values of the probe transmission and of the effective width $W$ are slightly smaller for the $x/y$ configuration than those for the $\sigma^+/\sigma^-$ configuration.

\begin{figure*}
\begin{center}
\includegraphics[width=7 cm]{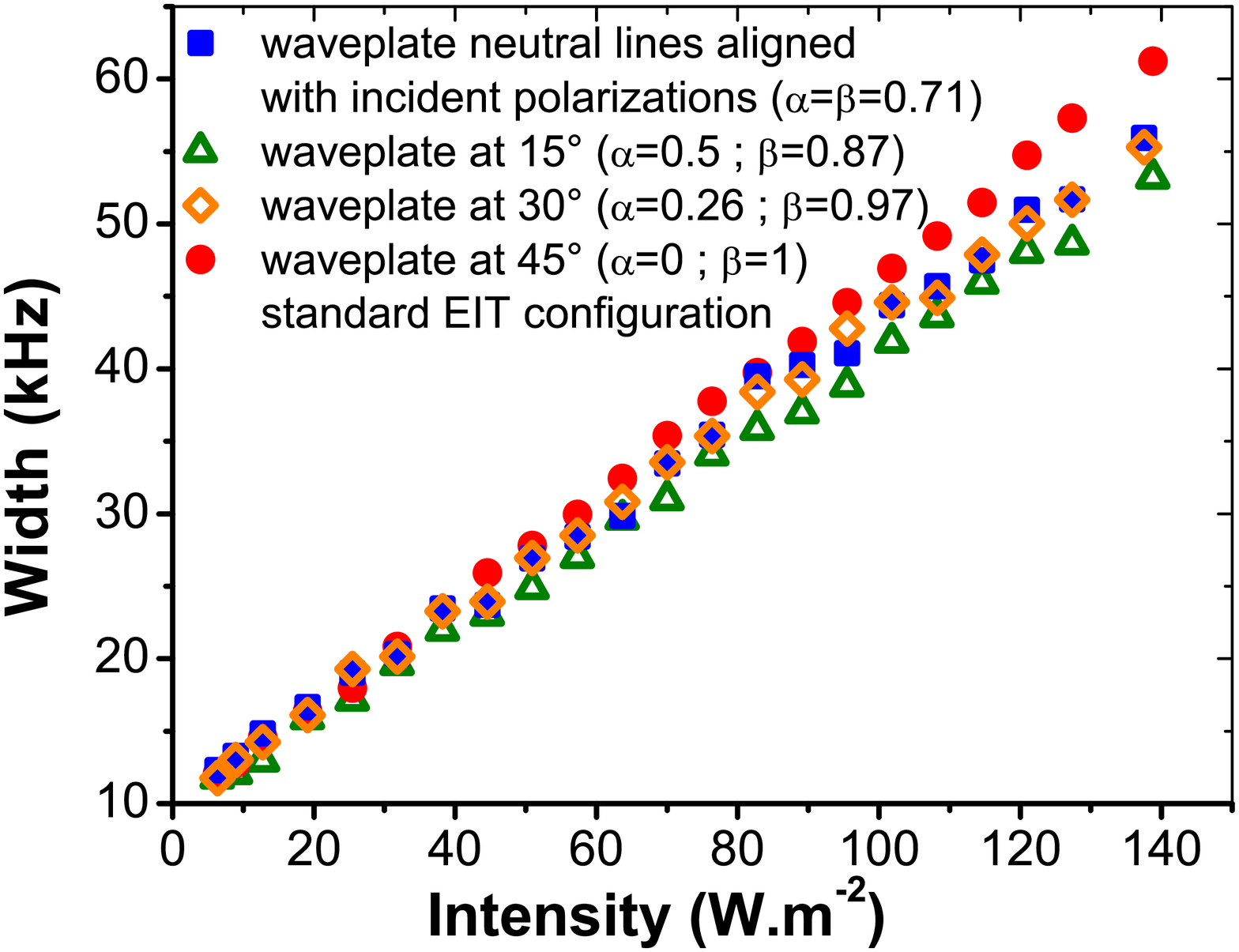} \hspace{2 cm}
\includegraphics[width=7 cm]{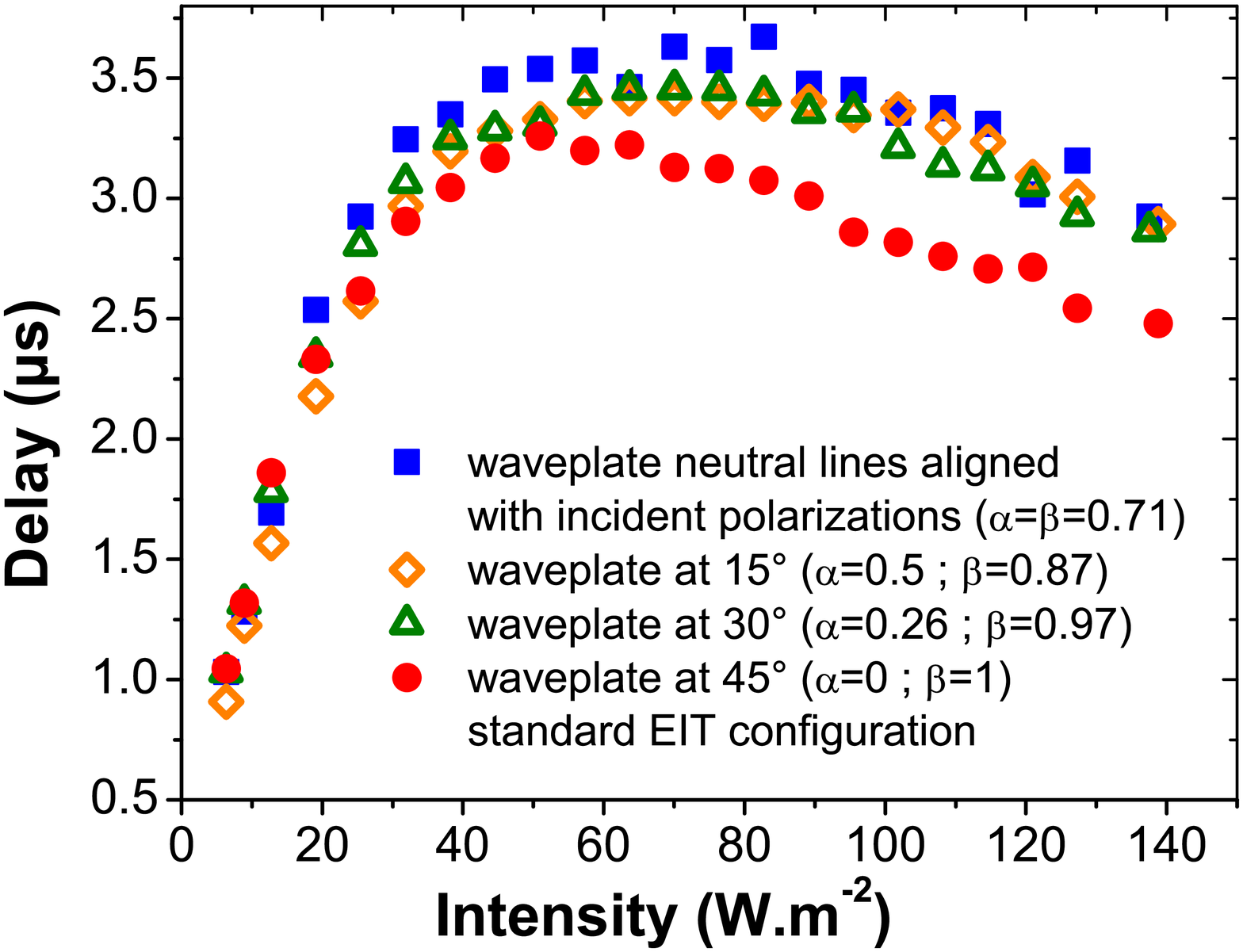}
\end{center}
\caption{Measured evolutions vs coupling beam intensity of the resonance width and group delay in standard EIT (circles) and symmetrical CPT (squares) configurations, and for $\alpha=0.87$ and $\beta=0.5$ (diamonds) and $\alpha=0.97$ and $\beta=0.26$ (triangles).}\label{Fig05}
\end{figure*}

One can wonder what would happen if the superposition of states was no longer symmetric. The mixing angle between the $|-1\rangle$ and $|+1\rangle$ sublevels in the new EIT basis can be chosen by adjusting the ellipticity of the polarizations of the coupling and probe beams. By choosing the proper orientation of the quarter-wave plate, we send in the helium cell two elliptic orthogonal polarizations, which make it possible to vary continuously the basis we are working with. The coupled and uncoupled states are labeled by $\alpha=\sin{(\pi/4-\theta)}$ and $\beta=\cos{(\pi/4-\theta)}$, where $\theta$ is the ellipticity of the coupling and probe beams:
\begin{subequations}
\label{eqTas4}
\begin{equation}
|C\rangle_\mathrm{C}=|NC\rangle_\mathrm{P}=\beta|-1\rangle-\alpha|+1\rangle\ ,\label{eq11}
\end{equation}
\begin{equation}
|NC\rangle_\mathrm{C}=|C\rangle_\mathrm{P}=\alpha|-1\rangle+\beta|+1\rangle\ ,\label{eq12}
\end{equation}
\end{subequations}
where $|C\rangle_\mathrm{C,P}$ and $|NC\rangle_\mathrm{C,P}$ are the coupled and non-coupled states corresponding to the coupling and probe laser beams. Figure \ref{Fig05} shows the behavior of width and delay with respect to the coupling beam averaged intensity for four different positions of the quarter-wave plate. These positions correspond to the standard $\sigma^+/\sigma^-$ EIT configuration ($\alpha=0$ and $\beta=1$), to the $x/y$ configuration of Fig.\,\ref{Fig01} ($\alpha=\beta=1/\sqrt{2}$) and to two intermediate sets of values for $\alpha$ and $\beta$.
We can see that the widths and delays are the same in all configurations. In the EIT case, the slope of the resonance width with respect to the intensity is slightly larger and the measured delays slightly smaller than in the other cases. This might be due to a residual magnetic field or a small difference in the effect of the velocity changing collisions. This will be further investigated in the future.

In conclusion, we have shown that EIT can be observed in three-level $\Lambda$ systems in which the two ground states are coherent superpositions of the eigenstates of the atomic total angular momentum given by the bright and dark states of CPT. In other words, any orthogonal superpositions of the ground-state sublevels are suitable for being the two feet of a $\Lambda$ system on which EIT can be built. Surprisingly, we have observed that the Raman coherence lifetime between these two states is the same as the Raman coherence lifetime between the usual $|m=-1\rangle$ and $|m=+1\rangle$ sublevels. Consequently, all the EIT features (transmission, EIT peak width, reduced group velocity for light) are the same in the two configurations. Such observations have been made possible in the case of the $^3$S$_1 \rightarrow ^3$P$_1$ transition of metastable helium because it constitutes a closed and perfect three-level system. This opens the way to interesting generalizations of the coherent processes in the so-called tripod system in the case of coherent superposition of the three ground-state sublevels \cite{Paspalakis2002,Vewinger2003}.

\begin{acknowledgments}
This work is supported by an Indo-French Networking Project funded by the Department of Science and Technology, Government of India, the French Ministry of Foreign Affairs and the Indo-French Centre for the Promotion of Advanced Research (IFCPAR/CEFIPRA). The authors are indebted to Herv\'e Gilles for providing them with some essential parts of the experimental set-up.
\end{acknowledgments}

\end{document}